\documentclass[reqno,12pt]{article}
\usepackage{amsmath,amssymb,amsfonts,epsfig,graphicx,euscript}
\usepackage{graphics}
\newcommand{\bse}{\begin{subequations}}
\newcommand{\ese}{\end{subequations}}
\newcommand{\be}{\begin{equation}}
\newcommand{\ee}{\end{equation}}
\newcommand{\bea}{\begin{eqnarray}}
\newcommand{\eea}{\end{eqnarray}}
\newcommand{\ba}{\begin{array}}
\newcommand{\ea}{\end{array}}

\begin{document}

\begin{flushright}
\end{flushright}
\begin{flushright}
\end{flushright}
\hfill%
\begin{center}

{\LARGE {\sc Hyperscaling-violating Lifshitz Solutions in Cubic
Gravity}}

\bigskip
{ Mohammad A. Ganjali
\footnote{ganjali@theory.ipm.ac.ir}} \\
{Department of Physics, Kharazmi University,\\P. O. Box
31979-37551, Tehran, Iran}
\\

\end{center}

\bigskip
\begin{center}
{\bf { Abstract}}\\
\end{center}
Considering the cubic theory of gravity which has been constructed
in \cite{myers}, we study the existence of Lifshitz and hyper
scaling violating Lifshitz solutions. We firstly extend the black
hole solution of \cite{myers} and find that such extended
solutions are valid for any value of dynamical exponent $z$.

Next, we examine the existence of the AdS black hole solution
with non-zero hyperscaling-violating exponent $\theta$ and
general dynamical exponent $z$. We find that the solutions do
exist only for $\theta=0,3$ in 4 dimension and $\theta=0,4$ in 5
dimension. In particular, when $\theta=3(4)$ in $4(5)$ dimension,
we have Lifshitz and Schwarzschild-AdS black hole solution
solution for $z=\{0,1\}$.

\newpage
\section{Introduction}
Recently, extended theories of gravity has attracted serious
attentions to study their roll in quantization of metric field,
black hole physics, AdS/CFT \cite{Maldacena:1997re}, cosmology
and etc. Several intelligent approaches have been constructed
such as extended theories with higher curvature, Horava-Lifshitz
gravity, bi-gravity, theories with higher spin fields, gravity
with higher dimensional effects and etc. See for example
\cite{Perez:2014pya}. Usually, the analysis of these theories are
based on Lagrangian formalism. Accordingly, One should firstly
find the Lagrangian of the fields presented in the theory.
Specially, for higher derivative theories, different
prescriptions was used for adding higher curvature terms to the
Einstein-Hilbert action. For example, in conformal gravity
\cite{Lu:2011zk}, one adds a Weyl tensor at quadratic order
$\int{W_{\mu\nu\lambda\theta}W^{\mu\nu\lambda\theta}d^4x}$. The
resultant action is an action which is invariant under the
conformal transformation.

Besides of some prior ideas and works on higher curvature gravity
\cite{lovel}, a beautiful way was recently suggested in
\cite{sinha} for finding higher order curvature action using a
concept which is called "Holographic c-Theorem". In fact,
considering a simple holographic c-theorem, the author of
\cite{sinha} was able to obtain constrains on new terms with
higher order in curvature. By generalizing this idea to higher
dimensions("Holographic a-theorem"), the action was calculated at
cubic and quartic order in curvature for arbitrary dimension
\cite{myers,Dehghani}. Also, another approach introduced for
calculating third order gravity in \cite{oliva} where it was used
the general properties of Wyle tensor. After finding these new
actions, it was studied various aspects of these new extended
theories of gravity such as its black hole solution and etc
\cite{myers,Dehghani}(see also \cite{Hennigar:2015esa} as some
recent works). There are some other approaches to study higher
curvature theories of gravity, see for example \cite{tekin}.

The problem of finding the solution of equation of motion for
complete cubic theory would be important and interesting. A class
of very important and attractive solutions are Lifshitz geometry.
The background metric of such geometry is give by
\cite{Kachru:2008yh}
 \bea\label{lifshitz}
ds^2=-\frac{r^{2z}}{L^2}dt^2+\frac{L^2}{r^2}dr^2+\frac{r^2}{L^2}d\vec{x}^2,
 \eea
where $z$ is dynamical exponent which exhibits space and time
scale differently in Lifshitz geometry. In fact, Such geometry
admits anisotropic scaling property
 \bea
t\mapsto \lambda^{z}t,\;\;\;\;r\mapsto
\lambda^{-1}r,\;\;\;\;\vec{x}\mapsto\lambda\vec{x}.
 \eea
A generalization of this geometry is conformally Lifshitz
geometry with the following line element \cite{Charmousis:2010zz}
 \bea\label{lifshitz}
ds^2=r^{-2\frac{\theta}{d-1}}\left(-\frac{r^{2z}}{L^2}dt^2+\frac{L^2}{r^2}dr^2+\frac{r^2}{L^2}d\vec{x}^2\right).
 \eea
Here $\theta$ is called the hyperscaling violation exponent. In
the context of AdS/CFT, the non-zero hyperscaling violation
exponent means that in dual field theory the hyperscaling
violates and the entropy scales as $T^{\frac{d-\theta-1}{z}}$.
These solutions and their properties have been studied in several
papers for the both Lifshitz and conformally Lifshitz geometries,
see for example \cite{AyonBeato:2010tm,mann}. It is noteworthy
that (conformally) Lifshitz geometries may also be generated in
theories where the gravity and matter coupled to each other
\cite{Taylor:2008tg}. Let us define $w=\frac{\theta}{d-1}$ in the
rest of the paper.

Considering the facts that a curvature cubed theory of gravity
which was obtained in \cite{myers} involves the standard Einstein
term, cosmological constant term, Gauss-Bone term and three
parameters family of curvature cubed terms potentially can
produce anisotropy in space and time, we attempt to find
(conformally)Lifshiz and asymptotically (conformally)Lifshiz-black
hole solution for such curvature cubed theory of gravity. We
shall firstly extend the black hole solution of \cite{myers} and
find that such extended solutions are valid for any value of
dynamical exponent $z$ and are also degenerate. We shall also
examine the existence of the Schwarzschild-AdS black hole solution
with non-zero hyper scaling violating exponent $\theta$. We will
find that the solutions do exist only for $w=0,1$. In particular,
when $w=1$, we have Lifshitz solution for $z=\{0,1,6\}$ and
Schwarzschild-AdS black hole solution for $z=\{0,1\}$ with certain
constrains on parameters of the theory.

The paper organized as follows. In the next section, we will
briefly introduce curvature cubed gravity. In section 3, we
consider the solution where was found in \cite{myers} for complete
cubic theory of gravity in 5 dimension. In section 4, we study
(conformally)Lifshiz and (conformally)Lifshitz-black hole
solution of cubic gravity in 4 and 5 dimensions. Finally, we end
up the paper by concluding remarks.
\section{Action of Cubic Gravity}
To begin, we firstly briefly review curvature cubed theory of
gravity which was introduced in \cite{myers}. The action of such
cubic order in curvature can be written as following
 \bea\label{action}
 I=\frac{1}{2l_p^{d-1}}\int\mathrm{d}^{d+1}x \, \sqrt{-g}\,
 \left(
\frac{d(d-1)}{L^2}\alpha+R+ L^2 {\cal X}+ L^4 {\cal Z}\right),
 \eea
where ${\cal X}$ and ${\cal Z}$ contain interactions at quadratic
and cubic order of curvature. Hereafter, we will consider $L=1$.
In \cite{myers,oliva}, it was shown that ${\cal X}$ is the
four-dimensional Euler density \cite{lovel},
 \bea
{\cal X}_4=R_{abcd}R^{abcd}-4R_{ab}R^{ab}+R^2
 \eea
For curvature-cubed interactions, it was also argued that a
three-parameter family of unitary $R^3$ interactions would exist.
It can be described in terms of a basis of three independent
interactions. The first of these is the cubic Lovelock
interaction which is proportional to the six-dimensional Euler
density ${\cal X}_6$ and is given by
 \bea
 {\cal X}_6&=& \frac{1}{8}\,
\varepsilon_{abcdef}\,\varepsilon^{ghijkl}\,R_{ab}{}^{gh}\,
R_{cd}{}^{ij}\, R_{ef}{}^{kl}\\ &=& 4\, R_{ab}^{\,\,\,\,\,\,cd}
R_{cd}^{\,\,\,\,\,\,ef} R_{ef}^{\,\,\,\,\,\, ab}-8\,
R_{a\,\,b}^{\,\,c\,\,\,d} R_{c\,\,d}^{\,\,e\,\,\,f}
R_{e\,\,f}^{\,\,a\,\,\,b} -24\, R_{a b c d} R^{a b
c}_{\,\,\,\,\,\,\,e} R^{d e}
 \nonumber\\
&&+3\, R_{a b c d} R^{a b c d} R+24\,R_{a b c d} R^{a c}R^{b
d}+16\, R_a^{\,\,b} R_b^{\,\,c} R_c^{\,\,a} -12\, R_a^{\,\,b}
R_b^{\,\,a} R + R^3\nonumber
 \eea
For $d=5$, it does not contribute to the equations of motion. For
$d\le 4$, this term vanishes.

The second basis interaction is the quasi-topological interaction
${\cal Z}_{d+1}$,
 \bea
 {\cal Z}_{d+1}&=& R_a{}^c{}_b{}^{d} R_c{}^e{}_d{}^{f}R_e{}^a{}_f{}^{b}
 +
\frac{1}{(2d-1)(d-3)}\left(\frac{3(3d-5)}{8}R_{a b c d}R^{a b c d}
R \right.\nonumber\\
&&\;\;\;\;\;\;-\,3(d-1) R_{a b c d}R^{a b c}{}_{e}R^{d e}+
3(d-1)R_{a b c d}
R^{a c}R^{b d}\\
&&\;\;\;\;\;\;\left.+\,6(d-1)R_a{}^{b}R_b{}^{c}R_c{}^{a}-\frac{3(3d-1)}{2}
R_a{}^{b}R_b{}^{a}R +\frac{3(d-1)}{8}R^3\right)\,.\nonumber
 \eea
 This term was only constructed for
$d=4$ and $d\ge6$. So, we don't have ${\cal Z}_{d+1}$ as a basis
interaction in $d\le3$ or $d=5$. Two other basis for the cubic
interaction are constructed from Weyl tensor
 \bea
{\cal W}_1=W_{a\,\,b}^{\,\,c\,\,\,d}\, W_{c\,\,d}^{\,\,e\,\,\,f}\,
W_{e\,\,f}^{\,\,a\,\,\,b}\,,\;\;\;\;\;\; {\cal
W}_2=W_{ab}^{\,\,\,\,\,\,cd}\, W_{cd}^{\,\,\,\,\,\,ef}\,
W_{ef}^{\,\,\,\,\,\, ab}\,.
 \eea
 where
  \bea
W_{abcd}=R_{abcd}-{2\over d-1} \left(
g_{a[c}\,R_{d]b}-g_{b[c}\,R_{d]a}
 \right)+{2\over d(d-1)}\,R\,g_{a[c}\,g_{d]b}
 \eea
 is the Weyl Tensor in $d+1$ dimensions\footnote{Here, it was used the standard notation $X_{[ab]}=\frac{1}{2}\left( X_{ab} -
X_{ba}\right)$.}. In fact, these terms do not change the
linearized equations of motion. But, as it was noted in
\cite{myers}, such interactions would effect some other
properties of the boundary QFT. For example, they would
contribute in calculation of the correlation function of the
stress tensor.

The above interactions are not all independent for $d\ge6$. In
fact, we have the following relation between the basis
interactions
 \bea
 {\cal Z}_{d+1}={\cal W}_1+\frac{3d^2-9d+4}{8(2d-1)(d-3)(d-4)}\left({\cal X}_6+8{\cal
 W}_1
 -4{\cal W}_2\right)\,.
 \eea
Hence, one can use any three of the above interactions as basis
for the curvature-cubed interactions when $d\ge6$. For $d=5$,
${\cal Z}_6$ is not defined and so the basis are ${\cal X}_6$,
${\cal W}_1$ and ${\cal W}_2$. For $d<5$, Schouten identities
reduce the number of possible interactions with ${\cal X}_6=0$
and ${\cal W}_1={\cal W}_2$ \cite{Sinha:2010pm}. Thus, for $d=4$,
we have a two parameter family of interactions with ${\cal Z}_5$
and ${\cal W}_1$. For $d=3$, ${\cal Z}_4$ is also not defined and
so we have a one parameter interaction with only ${\cal W}_1$ .

Henceforth in this paper, we focus our attention to $d=3$ and
$d=4$ dimensions. The generalization to higher dimensions is
straightforward. Our aim is to find static spherically symmetric
hyperscaling violated-Lifshitz solutions of full third order
action (\ref{action}). So, let us rewrite the action
(\ref{action}) as
 \bea\label{actionn}
I=\frac{1}{2l_p^{d-1}}\int\mathrm{d}^{d+1}x\sqrt{-g}\left(d(d-1)\alpha
+R+\lambda{\cal X}_4 +\mu {\cal Z}_{d+1}+\beta{\cal W}\right),
 \eea
where $\alpha$ is cosmological constant and $\lambda, \mu, \beta$
are three arbitrary real parameters\footnote{In literatures, the
coefficients $\lambda$ and $\mu$ usually are taken to be
$\lambda\rightarrow \frac{\lambda}{(d-2)(d-3)}$ and
$\mu\rightarrow -\frac{8(2d-1)\mu}{(d-5)(d-2)(3d^2-21d+4)}$.}and
${\cal W}={\cal W}_1={\cal W}_2$. We will also consider the
following static spherically symmetric ansatz,
 \bea\label{ansatz}
ds^2=-f^2(r)dt^2+\frac{dr^2}{h^2(r)}+j^2(r)dl^2_{k},
 \eea
where $dl^2_{k}$ is the line element for spherical, flat and
hyperboloid geometry as
 \bea
dl^2_k=dx_1^2+k^{-1}sin^2(\sqrt{k}x_1)\left(dx_2^2+\sum_{i=3}^{d-1}\prod_{j=2}^{i-1}sin^2\theta_jd\theta_i^2\right)
 \eea
For the above metric, one can easily show that all curvature
dependent functions can be rewritten using the following four
independent variables
 \bea
m&=&g^{rr}g^{ii}R_{riri},\;\;\;\;\;\;\;\;\;\;\;\;
n=g^{tt}g^{rr}R_{trtr}\cr
p&=&g^{tt}g^{ii}R_{titi},\;\;\;\;\;\;\;\;\;\;\;\;\;\;q=g^{ii}g^{jj}R_{ijij}.
 \eea
where indices $i$ stands for $x_i$ coordinates. In particular, one
can obtain
 \bea
 R&=&2n+2(d-1)m +2(d-1)p+(d-1)(d-2)q\\
{\cal X}_4&=&(d-1)(d-2)\left( 8 m p + 4nq+ \right. \nonumber\\
&&\hspace{1.5cm}\left. 4 (d-3) m q+4(d-3)qp+ (d-3)(d-4)q^2 \right)\\
 {\cal Z}_{d+1}&=&\frac{(d^2-3d+2) (3d^2-9d+4)}{8(2d-1)}\\ && \left(q \left(6 m \left(4 p + \left(d-5\right) q\right) +
   q \left(6 n + \left(d-5\right) \left(6 p + (d-6) q\right)\right)\right)\right)\nonumber\\
{\cal W}_1&=&{\cal W}_2=-\frac{8 (d-2) (d^3-6d^2+11d-4)}{
d^2(d-1)^2} \left(m-n+p-q\right)^3
 \eea
Evaluating the action (\ref{action}) for the ansatz (\ref{ansatz})
and varying it with respect to functions $f(r),h(r)$ and $j(r)$
gives us three equation of motion for these fields. It is worth
noting to recall that whenever $w=0$ we won't have hyperscaling
violation. Therefore, for spherically symmetric background
$j(r)=r$ and we would have two equations for $f(r)$ and $h(r)$.
The resulting equations of motion are very complicated and
tedious. In the next section, we will present reduced form of
them and try to find the hyper violating solutions.
\section{Hyper Violating Lifshitz Solution}
In this section, we will find the equation of motion and its
solutions for the full action of cubic gravity (\ref{actionn}).
For this aim, we perform the computations from two ways. Firstly,
in the next section, we generalize the black hole solution where
found in \cite{myers} for the action (\ref{actionn}) with
$\beta\neq 0$. Secondly, we attempt to find conditions for having
a simple hyper violating Lifshitz and Schwarzschild-AdS black hole
solution.
\subsection{Black Hole Solution} references
 As it was mentioned before, the author of \cite{myers} have found
 $5-$dimensional block hole solution using the action (\ref{actionn}) with
 $\alpha=1,\beta=0$ and the ansatz
  \bea
ds^2=-(k+r^2F(r))N(r)^2dt^2+\frac{dr^2}{(k+r^2F(r))}+r^2dl^2_{k}.
  \eea
  The solution is given by
  \bea\label{myerssol}
 &&N(r)=\frac{1}{F_{\infty}},\;\;\;\;F_{\infty}=\lim_{r\rightarrow \infty}F(r),\cr
 &&\alpha-F+2\lambda F^2+\frac{4}{7}\mu F^3=\frac{\omega^4}{r^4},
  \eea
where $\omega$ is an arbitrary constant. After the work of
\cite{myers}, the generalization to Lifshitz solution has been
done in \cite{mann}. In this case, using (\ref{ansatz}) with the
following replacements
 \bea\label{ansatzzz}
f^2(r)\rightarrow r^{2z}f(r),\;\;\;\;h^2(r)\rightarrow
r^2g(r),\;\;\;\;j(r)=r,
 \eea
one can again obtain the algebraic equation (\ref{myerssol}) for
$\kappa(r)$ where
 \bea\label{kappa}
\kappa(r)=g(r)-\frac{k}{r^2},
 \eea
 and two constraint on parameters
 \bea\label{myerssolpar}
\lambda=1-\frac{3}{2}\alpha,\;\;\;\;\;\;\mu=\frac{7}{4}(2\alpha-1).
 \eea
Interestingly, the above constraint and equation (\ref{myerssol})
are enough to solve the equation of motion and thus the function
$f(r)$ remains free and we have degenerate solution. It is easy
to see that if $\omega=0$ then $\kappa=1$ with the conditions
(\ref{myerssolpar}) would solve (\ref{myerssol}).

Now, we would like to examine whether such solution is a solution
of the full action (\ref{actionn}) with $\beta\neq 0$ or not. So,
it is enough to variate the last part of (\ref{actionn}) with
respect to $f(r)$ and $h(r)$ and find the equations which get from
this part. Due to third power of curvature, the resulting
equations are very complicated but, one can show that the
solutions of the following differential equation\footnote{In fact,
this differential equation comes from the factoring of equation
which is the result of variation of the action with respect to
$f(r)$.} do solve two equations of motion simultaneously,
 \bea
&&2f^2\left((z-1)r^2(rg'+2zg)-2k\right)\nonumber\\
&&\hspace{2cm}-r^4g'f^2+r^3f(f'\left(rg'+4zg)+2rgf''\right)=0.
 \eea
where the prime denotes the differentiation with respect to $r$.
The solution of the above equation reads as
 \bea\label{gg}
g(r)=\frac{\left(4kr^{2(z-1)}f(r)\pm
c^2\right)f(r)}{\left(rf'(r)+2(z-1)f(r)\right)^2}r^{-2z},
 \eea
where $c$ is an integration constant. As it is clear, up to this
stage, the function $f(r)$ again is free and we have degeneracy
on the solutions. The next step is to insert
$\kappa(r)=g(r)-\frac{k}{r^2}$ into (\ref{myerssol}) and solve
the new nonlinear equation. However, solving this differential
equation is very hard but for some simplified versions, one may
be able to solve it. In particular, we consider $\kappa=1$ which
means that $g(r)=1+\frac{k}{r^2}$. We also consider $\omega=0$.
Then, one can show that
 \bea
f(r)&=&\frac{1}{4}(c-2r\hat{c})^2r^{-2 z} ,\;\;\;\;\;\;\;\;\;\;\;\;k=0,\;\;c>0\\
f(r)&=&(2+ r^2\pm 2\sqrt{1+r^2]})r^{-2z},\;\;\;\;\;\;k=1,\;\;c=0.
 \eea
Note that in the above solutions, for $k=0$ the $c>0$ is necessary
but for $k=1$  we set zero all the integration constants for
simplicity. One can also find solution for $k=-1$ but the solution
is not a real function\footnote{For the specific case where
$f(r)=g(r)$, firstly, we solve (\ref{gg}). The result is
 \bea\label{myerssolsol}
&&f(r)=g(r)=\left(-c(z-2)^2r^2+4(r^{z-2}+\tilde{c}(z-2))^2\right)\frac{kr^{-2z}}{4(z-2)^2},\;\;\;k\neq
0,\cr
&&f(r)=g(r)=\left(\pm\sqrt{c}r^z+\tilde{c}(z-2)r^2\right)\frac{r^{-2z}}{(z-2)},\;\;\;k=0
 \eea
where $c, \tilde{c}$ are arbitrary constants. The above solutions
are singular at $z=2$. Inserting $z=2$ in (\ref{gg}) and solving
it with the condition $f(r)=g(r)$ one obtains
 \bea
f(r)=g(r)=\frac{1}{4kr^2}\left(-c+4k^2\tilde{c}^2-8\tilde{c}k^2\log{r}+4k^2\log{r}^2\right)
 \eea
 which is a \textit{log} solution with singularity in $r=0$. Then, we should insert these solutions into
 (\ref{myerssol}) to find constraint on coefficients.}.

As a summary, the solution (\ref{gg}) with
(\ref{myerssol}-\ref{myerssolpar}) is a complete solution of
equations of motion for the full action (\ref{actionn}) with
arbitrary $\beta$.
\subsection{Schwarzschild-AdS Solution}
In this section, our aim is to obtain the simple Schwarzschild-AdS
solution. In fact, we want to find hyper scaling violating
Lifshitz solution \cite{Charmousis:2010zz,AyonBeato:2010tm} by
using the following redefinitions in the ansatz (\ref{ansatz}) ,
 \bea\label{ansatzz}
f(r)=r^{z-w-1}g(r),\;\;\;\;h(r)=r^{w}g(r),\;\;\;\;j(r)=r^{1-w}.
 \eea
In this section, we also perform more simplification by
considering
 \bea\label{g}
g(r)=r^2-s
 \eea
with some arbitrary constant $s$. By these assumptions, we search
the conditions on parameters $\alpha, \lambda, \mu, \beta$ and
$s$ in which $h(r)=r^{w}(r^2-s)$, $f(r)=r^{z-w-1}(r^2-s)$ would
be the solutions of equation of motion. We will do the
computations for $4$ and $5$-dimensional cases separately. The
generalization to higher dimensions is straightforward.
\subsubsection{4-Dimension}
First of all, Recall that in 4 dimension, ${\cal W}_1={\cal W}_2$
and ${\cal X }_4$ is a topological term. Then, varying the action
(\ref{actionn}), evaluated for the ansatz (\ref{ansatz}), with
respect to $f(r), h(r)$ and $j(r)$ and inserting (\ref{ansatzz})
and (\ref{g}) into the equations of motion, one finds that the
following polynomials should be zero for all range of $r$,
 \bea\label{poly}
P^{h,f,j}(r)=\sum_{a,b=0}^{3}P^{h,f,j}_{ab}r^{2a+2bw}=0,
 \eea
where $P^{h,f,j}_{ab}$ are some constants and are given
by\footnote{Here, $h,f$ and $j$ stands for variation of the action
with respect to $h(r)$ and $f(r)$ respectively.}
 \bea
P^h_{00}&=&-4 \beta k^3,\\
P^h_{10}&=&0,\nonumber\\
P^h_{20}&=&-18k,\nonumber\\
P^h_{30}&=&-54\alpha,\nonumber\\
P^h_{01}&=&24 \beta k^2 s (1 + w) (-2 + z),\nonumber\\
P^h_{11}&=&-24 \beta k^2 (1 + w) (-1 + z),\nonumber\\
P^h_{21}&=&-18 s (-1 + w) (1 + 3 w - 2 z),\nonumber\\
P^h_{31}&=&18 (-1 + w) (-1 + 3 w - 2 z),\nonumber\\
P^h_{02}&=&12 \beta k s^2 (-2 + z)^3 (2 + z),\nonumber\\
P^h_{12}&=&-24 \beta k s (-2 + z)^2 (-2 + z + z^2),\nonumber\cr
P^h_{22}&=&12 \beta k (-1 + z)^2 z^2,\nonumber\\
P^h_{32}&=&0,\nonumber\\
P^h_{03}&=&-8 \beta s^3 (-1 + 3 w - z) (-2 + z)^5,\nonumber\\
P^h_{13}&=&24 \beta s^2 (-2 + z)^3 (-1 + z) (2 + z - z^2 + w (-2 + 3 z)),\nonumber\\
P^h_{23}&=&24 \beta s (-2 + z) (-1 + z)^2 z (4 w - 3 (1 + w) z + z^2),\nonumber\\
P^h_{33}&=&8 \beta (3 + 3 w - z) (-1 + z)^3 z^2,\nonumber
 \eea
 and
 \bea
\hspace{1cm}P^f_{00}&=&-4 \beta k^3,\\
P^f_{10}&=&0,\nonumber\\
P^f_{20}&=&-18 k,\nonumber\\
P^f_{30}&=&-54\alpha,\nonumber\\
P^f_{01}&=&-48 \beta k^2 s (1 + w)^2,\nonumber\\
P^f_{11}&=&24 \beta k^2 (1 + w) (1 + 2 w),\nonumber\\
P^f_{21}&=&-18 s (-1 + w)^2,\nonumber\\
P^f_{31}&=&18 (-3 + w) (-1 + w),\nonumber\\
P^f_{02}&=&12 \beta k s^2 (-2 + z)^2 (-4 + (-4 + z) z),\nonumber\\
P^f_{12}&=&-24 \beta k s (-2 + z)^3 (1 + z),\nonumber\\
P^f_{22}&=&12 \beta k (-1 + z)^2 z^2,\nonumber\\
P^f_{32}&=&0,\nonumber\\
P^f_{03}&=&8 \beta s^3 (-2 + z)^4 (-2 - 6 (-2 + w) w + (-4 + z) z),\nonumber\\
P^f_{13}&=&-24 \beta s^2 (-2 + z)^2 (-4 + 3 w (-2 + z)^2 + z^2 (7
+
(-5 + z) z)\nonumber\\
   &&-2 w^2 (4 + 3 (-2 + z) z)),\nonumber\\
P^f_{23}&=&24 \beta s (-1 + z) z ((-1 + z) z (2 + (-4 + z)
z)\nonumber\\ && -
   2 w^2 (8 + 3 (-3 + z) z) - 2 w (4 + 3 (-2 + z) z)),\nonumber\\
P^f_{33}&=&8 \beta (-1 + z)^2 z^2 (9 + 3 w (5 + 2 w) + z -
z^2),\nonumber
 \eea
and
 \bea
\hspace{1cm}
P^j_{00}&=&4 \beta k^3,\\
P^j_{10}&=&0,\nonumber\\
P^j_{20}&=&18 k,\nonumber\\
P^j_{30}&=&54\alpha,\nonumber\\
P^j_{01}&=&-12 \beta k^2 s (3 + w - z) (2 w + z),\nonumber\\
P^j_{11}&=&12 \beta k^2 (2 (1 + w)^2 - w z - z^2),\nonumber\\
P^j_{21}&=&18 s (1 + w - z)^2,\nonumber\\
P^j_{31}&=&-18 (1 + w^2 + z + z^2 - 2 w (1 + z)),\nonumber\\
P^j_{02}&=&12 \beta k s^2 (-2 + z)^2 (-4 + (-2 + z) z),\nonumber\\
P^j_{12}&=&-24 \beta k s (-2 + z)^2 (-2 + z^2),\nonumber\\
P^j_{22}&=&12 \beta k (-1 + z)^2 z^2,\nonumber\\
P^j_{32}&=&0,\nonumber\\
P^j_{03}&=&-4 \beta s^3 (-2 + z)^4 (8 + 6 (-3 + w) w + z + 3 w z - z^2),\nonumber\\
P^j_{13}&=&-12 \beta s^2 (-2 +
   z)^2 (-2 w^2 (4 + 3 (-2 + z) z) +\nonumber\\
   && (-2 + z)^2 (-2 + z^2)
   -w (-2 + z) (8 + z (-8 + 3 z))),\nonumber\\
P^j_{23}&=&12 \beta s (-1 + z) z ((-4 + z) (-1 + z)^2 z
+\nonumber\\&& w z (-6 + (7 - 3 z) z) -
   2 w^2 (8 + 3 (-3 + z) z)),\nonumber\\
P^j_{33}&=&4 \beta (-1 + z)^2 z^2 (6 + 6 w^2 - (-4 + z) z + 3 w (4
+ z)),\nonumber
 \eea
Now, for finding the solutions, a few comments are in order.
Firstly, we would like to find the solution by imposing the
condition $\lambda\neq 0$ \textit{and} $\mu\neq 0$ \textit{and}
$\beta\neq 0$. Secondly, as far as one choose an specific value
for $w$, then some powers of $r$ in (\ref{poly}) would be equal to
each other and so the sum of the coefficients of these terms
should be set to zero. For example, if $w=1$ then the power of
$r$ for $a=0,b=3$ and $a=3,b=0$ are equal and so we should set
the sum of $P_{03}$ and $P_{30}$ to zero. Finally, noting the
coefficients of $P_{31}^{f,h}$ and $P_{33}^{f,h}$ implies that
 \bea
 -3\leq w\leq 3.
 \eea
By further analysis, one can show the solution with the condition
$\lambda\neq 0$ \textit{and} $\mu\neq 0$ \textit{and} $\beta\neq
0$ do exist only for the following two cases
 \bea
 w=0,1.
 \eea
The case $w=0$ is the usual geometry without scaling violation
but for $w=1$ we have hyper scaling violation. In $4$ dimension,
the $w=1$ case implies that hyper scaling violation exponent
$\theta$ should be equal to $3$.
\begin{itemize}
\item{$w=0$}\\
In this case, the solution depends on the value of $z$.
 \bea
a)&&\;\;\;\;s=0,\;\;\;\;k=0,\;\;\;\;\alpha\neq
0,\;\;\;\;z\neq \{0,1,4\}\hspace{3cm}\nonumber\\
&&\;\;\;\; \alpha=\frac{18 + 7 z + 4 z^2 - 2 z^3}{9(4 -
z)},\;\;\;\beta=-\frac{3}{2 z^2 (4 - 5 z + z^2)},\nonumber\\
 b)&&\;\;\;\;s=-k,\;\;\;\;\alpha\neq
0,\;\;\;\;z=\{1\}\nonumber\\
&&\;\;\;\; \alpha=1,\;\;\;\;\;\;\;\;\beta=arbitrary.
 \eea
The case $a$ is the standard Lifshitz geometry (\ref{lifshitz}).
 The second case with $s=-k$ gives us the black hole solution but
 only for $z=1$ and arbitrary $\beta$.

 Here, there is not consistent solution for
$z=\{0,4\}$\footnote{Note that we always consider $\beta\neq 0.$}.
\item{$w=1$}\\
The solution depends on $z$ and  whether $s$ is equal to zero or
not.
 \bea
a)&&\;\;\;\;s=0,\;\;\;\;k=0,\;\;\;\;\alpha=0\;\;\;\;z=\{0,1\},\hspace{3cm}\nonumber\\
&&\;\;\;\;\beta=arbitrary\\
b)&&\;\;\;\;s\neq 0,\;\;\;\;k=0,\;\;\;\;\alpha\neq
0\;\;\;\;z=\{0,1\}\nonumber\\
&&\;\;\;\;s=-\frac{(2z+1)^2}{(z+1)}\alpha,\;\;\;\;\;\;\beta=\frac{27}{4(z-2)^6}\frac{\alpha}{s^3}.\nonumber
 \eea
 The case $a$ again gives us the Lifshitz geometry (\ref{lifshitz})
 but only for $z=\{0,1\}$. The second case $b$ gives us a black
 hole solution but only with flat horizon.
\end{itemize}
\subsubsection{$5$-Dimension}
In $5$-dimension, we should also consider the ${\cal X}_4$ and
${\cal Z}_5$ terms. The remaining procedure is quite similar to
$4$ dimensional. Using the ansatz (\ref{ansatzz}) and (\ref{g}) in
equations of motion for $f(r), h(r)$ and $j(r)$, one finds the
coefficients $P^{f,h,j}_{ab}$ as follows
 \bea\hspace{1.1cm}
P^h_{00}&=&8 (27 \mu + 7 \beta) k^3\\
P^h_{10}&=&0,\nonumber\\
P^h_{20}&=&378 k,\nonumber\\
P^h_{30}&=&756\alpha,\nonumber\\
P^h_{01}&=&-24 (27 \mu + 7 \beta) k^2 s (1 + 3 w) (-2 + z),\nonumber\\
P^h_{11}&=&24 k (27 \mu k (1 + 3 w) (-1 + z) + 7 \beta k (1 + 3 w)
(-1 + z) \nonumber\\ &&+
   63 \lambda s (-1 + 5 w^2 + z - 3 w z)),\nonumber\\
P^h_{21}&=&-378 (- s (-1 + w) (2 w - z) + 4 \lambda k (5 w^2 + z - 3 w (1 + z))),\nonumber\\
P^h_{31}&=&-378 (-1 + w) (-1 + 2 w - z),\nonumber\\
P^h_{02}&=&24 k s^2 (27 \mu (-1 + w)^2 (3 + w (14 + 7 w - 10 z) -
2 z) \nonumber\\ &&-
   7 \beta (-2 + z)^3 (2 w + z)),\nonumber\\
P^h_{12}&=&24 s (-54 \mu k (-1 + w)^2 (2 + w (9 + 7 w - 10 z) - 2
z) \nonumber\\ &&+
   7 (9 \lambda s (-1 + w)^3 (1 + w - z)\nonumber\\ && +
      2 \beta k (-2 + z) (-1 + z) (-2 + 2 w (-1 + z) + (-2 + z) z))),\nonumber\\
P^h_{22}&=&24 (27 \mu k (-1 + w)^2 (1 + w (4 + 7 w - 10 z) - 2 z)
\nonumber\\ &&-
   7 (9 \lambda s (-1 + w)^3 (1 + 2 w - 2 z) +
      \beta k (-1 + z)^2 z (-2 + 2 w + z))),\nonumber\\
P^h_{32}&=&1512 \lambda (-1 + w)^3 (w - z),\nonumber\\
P^h_{03}&=&-8 s^3 (27 \mu (-1 + w)^5 (4 + 2 w - 3 z) -
   7 \beta (1 + 3 w - 2 z) (-2 + z)^5),\nonumber\\
P^h_{13}&=&24 s^2 (27 \mu (-1 + w)^5 (3 + 2 w - 3 z)\nonumber\\ &&
+
   7 \beta (-2 + z)^3 (-1 + z) (-2 + w (2 - 3 z) + z (-5 + 2 z))),\nonumber\\
P^h_{23}&=&-24 s (27 \mu (-1 + w)^5 (2 + 2 w - 3 z)\nonumber\\ &&
+
   7 \beta (-2 + z) (-1 + z)^2 z (4 (1 + w) - 3 (3 + w) z + 2 z^2)),\nonumber\\
P^h_{33}&=&8 (27 \mu (-1 + w)^5 (1 + 2 w - 3 z) -
   7 \beta (9 + 3 w - 2 z) (-1 + z)^3 z^2),\nonumber
 \eea
 and
 \bea
P^f_{00}&=&8 (27 \mu + 7 \beta) k^3,\\
P^f_{10}&=&0,\nonumber\\
P^f_{20}&=&378k,\nonumber\\
P^f_{30}&=&756\alpha,\nonumber\\
P^f_{01}&=&24 (27 \mu + 7 \beta) s (k + 3 k w)^2,\nonumber\cr
\hspace{.8cm}P^f_{11}&=&-72 k w (27 \mu (k + 3 k w) + 7 (-12
\lambda s w +
\beta (k + 3 k w))),\nonumber\\
P^f_{21}&=&-378 (-s (-1 + w)^2 + 4 \lambda k (1 + w (-3 + 4 w))),\nonumber\\
P^f_{31}&=&-378 (-2 + w) (-1 + w),\nonumber\\
P^f_{02}&=&24 k s^2 (27 \mu (-1 + w)^2 (1 + w (6 + w))\nonumber\\
&& -
   7 \beta (-2 + z)^2 (2 - 2 w (2 + w) + (-4 + z) z)),\nonumber\\
P^f_{12}&=&-48 k s (27 \mu (-1 + w)^2 w (1 + w) \nonumber\\ &&+
   7 \beta (-(-3 + z) (-1 + z)^2 z - w (-4 + z (2 + z)) \nonumber\\ &&+
      w^2 (4 + z (-5 + 2 z)))),\nonumber\\
P^f_{22}&=&24 (27 \mu k (-1 + w)^2 (-1 + (-4 + w) w) +
   7 (9 \lambda s (-1 + w)^3 \nonumber\\ &&-
      \beta k (-1 + z) z (-4 - 2 (-3 + w) w + (-1 + z) z))),\nonumber\\
P^f_{32}&=&-1512 \lambda (-1 + w)^3,\nonumber\\
P^f_{03}&=&8 s^3 (27 \mu (-1 + w)^6\nonumber\\ && -
   7 \beta (-2 + z)^4 (5 - 3 (-2 + w) w + 2 (-4 + z) z)),\nonumber\\
P^f_{13}&=&-24 s^2 (27 \mu (-1 + w)^5 w \nonumber\\ &&+
   7 \beta (-2 + z)^2 (-2 (-3 + z) (-1 + z)^2 z \nonumber\\ &&+ w^2 (4 + 3 (-2 + z) z) +
       w (-4 + 3 z^2))),\nonumber\\
P^f_{23}&=&24 s (27 \mu (-1 + w)^5 (1 + w) \nonumber\\ &&-
   7 \beta (-1 + z) z (-16 + w^2 (-8 - 3 (-3 + z) z) \nonumber\\ &&-
      6 w (4 + z (-5 + 2 z)) + z (17 + z (3 + 2 (-5 + z) z)))),\nonumber\\
P^f_{33}&=&-8 (27 \mu (-1 + w)^5 (2 + w)\nonumber\\ && +
   7 \beta (-1 + z)^2 z^2 (36 + 3 w (7 + w) - 2 (-1 + z) z)),\nonumber
 \eea
 and
 \bea\hspace{1.5cm}
P^j_{00}&=&8 (27 \mu + 7 \beta) k^3,\\
P^j_{10}&=&0,\nonumber\\
P^j_{20}&=&378 k,\nonumber\\
P^j_{30}&=&756\alpha,\nonumber\\
P^j_{01}&=&-8 (27 \mu + 7 \beta) k^2 s (-9 + 9 w^2 - 3 w (-4 + z) + (11 - 3 z) z),\nonumber\\
P^j_{11}&=&8 k (27 \mu k (1 + 9 w^2 - 3 w (-2 + z) + (2 - 3 z) z)
+\nonumber\\ &&
   7 (\beta k (1 + 9 w^2 - 3 w (-2 + z) + (2 - 3 z) z) +\nonumber\\ &&
      9 \lambda s (4 + 2 w (5 + 3 w) - 7 z - 7 w z + 3
      z^2))),\nonumber\cr
P^j_{21}&=&-126 ( s (-1 - 3 w^2 + z + 3 w z - z^2) +\nonumber\\ &&
   4 \lambda k (2 + 6 w^2 + z (2 + 3 z) - w (4 + 7 z))),\nonumber\\
P^j_{31}&=&-126 (3 + 3 w^2 - 3 w (2 + z) + z (2 + z)),\nonumber\\
P^j_{02}&=&-8 k s^2 (7 \beta (-2 + z)^2 (-6 + 2 w (4 + w) + 10 z -
2 w z - 3 z^2) +\nonumber\\ &&
   27 \mu (-1 + w) (3 + w (-45 + (-23 + w) w) -\nonumber\\ && 6 z + 6 w (8 + w) z +
      2 (1 - 5 w) z^2)),\nonumber\\
P^j_{12}&=&8 s (54 \mu k (-1 + w) (-1 + w (-13 + (-11 + w) w)
-\nonumber\\ &&  z +
      w (23 + 6 w) z + 2 (1 - 5 w) z^2) +\nonumber\\ &&
   7 (-9 \lambda s (-1 + w)^2 (1 + w - z) (-2 + z) +\nonumber\\ &&
      2 \beta k (4 + 5 z - 3 z^2 (7 + (-5 + z) z) +
         w^2 (4 + z (-5 + 2 z)) +\nonumber\\ &&  w (8 + z (-12 + (7 - 2 z) z))))),\nonumber\\
P^j_{22}&=&-8 (7 z (-9 \lambda s (-1 + w)^2 (1 + 2 w - 2 z)
+\nonumber\\ &&
      \beta k (-1 + z) (2 + 2 w^2 + (5 - 3 z) z - 2 w (2 + z))) +\nonumber\\ &&
   27 \mu k (-1 + w) (-1 + w^3 + 2 z (2 + z) + \nonumber\\ && w^2 (1 + 6 z) -
      w (1 + 2 z (1 + 5 z)))),\nonumber\\
P^j_{32}&=&-504 \lambda (-1 + w)^2 (w - z) (2 + z),\nonumber\\
P^j_{03}&=&8 s^3 (27 \mu (-1 + w)^4 (-5 + w^2 - w z - (-5 + z) z)
+\nonumber\\ &&
   7 \beta (-2 + z)^4 (1 + 4 w - w^2 - (3 + w) z + z^2)),\nonumber\\
P^j_{13}&=&-8 s^2 (27 \mu (-1 + w)^4 (-5 + 3 w^2 + w (2 - 3 z) +
(8 - 3 z) z) -\nonumber\\ &&
   7 \beta (-2 + z)^2 (4 - (-3 + z) (-2 + z) z (-1 + 3 z) +\nonumber\\ &&
      w^2 (4 + 3 (-2 + z) z) + w (-8 + z (12 + z (-8 + 3 z))))),\nonumber\\
P^j_{23}&=&8 s (27 \mu (-1 + w)^4 (1 + w (4 + 3 w) + z - 3 w z - 3
z^2) +\nonumber\\ &&
   7 \beta (-1 + z) z (-8 (1 + w)^2 +
      3 (-3 + w (4 + 3 w)) z + \nonumber\\ && (30 + w - 3 w^2) z^2 - (20 +
         3 w) z^3 + 3 z^4)),\nonumber\\
P^j_{33}&=&-8 (27 \mu (-1 + w)^4 ((1 + w)^2 - (2 + w) z - z^2)
-\nonumber\\ &&
   7 \beta (-1 + z)^2 z^2 ((3 + w)^2 + (4 + w) z - z^2)),\nonumber
 \eea
In this case, analyzing the coefficients $P^{f,h}_{31}$ and
$P^{f,h}_{32}$ implies that the allowed values for $w$ are
 \bea
w=0,1.
 \eea
So, the hyper scaling violation exponent is equal to $4$ and the
solutions are given by
\begin{itemize}
\item{$w=0$}\\
 \bea
a)&&\;\;\;\;s=0,\;\;\;\;k=0,\;\;\;\;z\neq\{0,1\}\hspace{6cm}\nonumber\\
&&\;\;\;\;\lambda=\frac{-3\alpha (1 + 15 z^2 - 18 z^3 + 3
z^4)}{2(1 + 9 z^2 + 16 z^3 - 21 z^4 - 6 z^5 + 2 z^6)}\nonumber
\\&&\;\;\;\;\;\;\;\;\;\; +\frac{(2 + 27 z^2 - 19 z^3 - 6 z^4 - 3 z^5 + z^6)}{2(1 + 9
z^2 + 16 z^3 -
21 z^4 - 6 z^5 + 2 z^6)}\nonumber \\
&&\;\;\;\;\mu=\frac{7(2\alpha-1)}{4(1 + 9 z^2 + 16 z^3 - 21 z^4 -
6 z^5 +
2 z^6)}\nonumber \\
&& \;\;\;\;\beta=\frac{27(2\alpha-1)}{4(1 + 9 z^2 + 16 z^3 - 21
z^4 - 6 z^5 + 2 z^6)}\\
b)&&\;\;\;\;s=0,\;\;\;\;k=0,\;\;\;\;z=\{0\}\nonumber\\
&&\;\;\;\;\lambda=1 -
\frac{3\alpha}{2},\;\;\;\;\mu=\frac{7}{4}(2\alpha-1),\;\;\;\;\beta=arbitrary,\\
c)&&\;\;\;\;s=0,\;\;\;\;k=0,\;\;\;\;z=\{1\}\nonumber\\
&&\;\;\;\;\lambda=\frac{1-\alpha}{2}-\frac{2\mu}{7},\;\;\;\;\;\;\mu,\beta=arbitrary,\\
d)&&\;\;\;\;s=-k,\;\;\;\;s,k\neq 0,\;\;\;\;z=\{1\}\nonumber\\
&&\;\;\;\;\lambda=\frac{1-\alpha}{2}-\frac{2\mu}{7},\;\;\;\;\;\;\mu,\beta=arbitrary,\\
e)&&\;\;\;\;s\neq k,\;\;\;\;s,k\neq 0\;\;\;\;\;\;z=\{1\}\nonumber\\
&&\;\;\;\;\lambda=1 -
\frac{3\alpha}{2},\;\;\;\;\mu=\frac{7}{4}(2\alpha-1),\;\;\;\;\beta=\frac{27}{4}(2\alpha-1)\\
f)&&\;\;\;\;s=-\frac{k}{4},\;\;\;\;s,k\neq 0,\;\;\;\;\;\;z=\{3\}\nonumber\\
&&\;\;\;\;\lambda=
\frac{1}{218},\;\;\;\;\mu=-\frac{7}{1308},\;\;\;\;\beta=\frac{9}{436}
 \eea
 The solutions $a,b,c$ are the Lifshitz geometry (\ref{lifshitz}).
 Thus in $5$ dimension, we have Lifshitz solution in cubic gravity for any value of
 $z$. Notice that the case (c) is in agreement with (\ref{myerssolsol}).
 The cases $d,e,f$ demonstrate black hole solution with
 various horizon geometry. These black hole solutions exist only
 for $z=1$ and $z=3$. Note also that the parameters $\lambda,\mu$
 and $\beta$ are exactly fixed for $z=3$.
\item{$w=1$}
 \bea\label{constraint5}
a)&&\;\;\;\;s=0,\;\;\;\;k=0,\;\;\;\;\alpha=0\;\;\;\;z=\{0,1\}\hspace{3cm}\nonumber\\
&&\;\;\;\;\lambda,\mu,\beta=arbitrary,\\
b)&&\;\;\;\;s\neq 0,\;\;\;\;k=0,\;\;\;\;\alpha\neq0\;\;\;\;z=\{0,1\}\hspace{3cm}\nonumber\\
&&\;\;\;\;s=-\frac{2(2z+1)}{z+1}\alpha\;\;\;\;\beta=\frac{27}{4\beta(z-2)^6}\frac{\alpha}{s^3},\nonumber\\
&&\;\;\;\lambda,\mu=arbitrary.
 \eea
The hyper scaling violating solutions in 5 dimension also contain
Lifshitz background in case $a$ but with flat boundary $k=0$ and
exist only for $z=\{0,1\}$. We also have black hole solution in
case $b$. But these solutions do also exist only for $z=\{0,1\}$
and we have the constraint (b) on parameters of the action
(\ref{actionn}).
\end{itemize}
\section{Conclusion}
In this paper, we study the existence of asymptotically Lifshitz
and hyper scaling violating asymptotically Lifshitz solutions in
full cubic theory of gravity in 4 and 5 dimensions with the action
(\ref{actionn}). Such cubic action of curvature has been
constructed in \cite{myers} using some simple "Holographic c/a
theorem" in arbitrary dimensions.

We firstly extend the black hole solution of \cite{myers} for full
cubic action(\ref{actionn}). This solution is valid for any value
of dynamical exponent $z$.

Next, we examine the usual Schwarzschild-AdS solution with
non-zero hyper scaling violating exponent $-2\frac{\theta}{d-1}$
and general dynamical exponent $z$. We have found that the
solutions do exist only for $\theta=0,3$ in 4 dimension and
$\theta=0,4$ in 5 dimension. In particular, when $\theta=0$, we
have Lifshitz solution for any value of $z$ except of
$z=\{0,4\}$(in 4 dimension). We also have Schwarzschild-AdS black
hole solution with $z=\{1\}1$ in 4 dimension and with $z=\{1,3\}$
in 5 dimension. The above solutions exist with certain constrains
on parameters of the theory.

Moreover, when $\theta=3(4)$ in $4(5)$ dimension, we have
Lifshitz and Schwarzschild-AdS black hole solution for
$z=\{0,1\}$.

At the end, we would like to mention that there are various fields
of study for theories with cubic action (\ref{actionn}) and their
solutions. For example, correlation functions in cubic gravity,
the thermodynamics of black hole solutions, the Noether charges
and conserved currents of solutions, their CFT dual and many other
problems which are open, important and interesting.
\section{Acknowledgment}
Mohammad A. Ganjali would like to thanks the Kharazmi university
for supporting the paper via grant.

\end{document}